%% file: main.tex
\documentclass{egpubl}
\usepackage{eurovis2025}

\EuroVisEducation              

\usepackage[T1]{fontenc}
\usepackage{dfadobe}

\usepackage{cite}  
\BibtexOrBiblatex
\electronicVersion
\PrintedOrElectronic
\ifpdf \usepackage[pdftex]{graphicx} \pdfcompresslevel=9
\else \usepackage[dvips]{graphicx} \fi

\usepackage{egweblnk}

\newcommand{\etal}{et~al.~} 
\newcommand{\ie}{i.e.,~}
\newcommand{\eg}{e.g.,~}

\usepackage{graphicx}
\usepackage{wrapfig}
\usepackage[T1]{fontenc}
\usepackage{amsmath}  
\usepackage{multirow}
\usepackage{tabularx}
\usepackage[svgnames,x11names,table]{xcolor}
\usepackage{soul}
\usepackage{xurl}

\title[Reflections on Teaching Data Storytelling at the Journalism School]%
      {Reflections on Teaching Data-Driven Storytelling \\ at the Journalism School}

\author[Lan]
{\parbox{\textwidth}{\centering Xingyu Lan\orcid{0000-0001-7331-2433}
        }
        \\
{\parbox{\textwidth}{\centering Fudan University, Shanghai, China\\
       }
}
}

\begin{document}

\teaser{
\vspace{-2em}
 \includegraphics[width=0.8\linewidth]{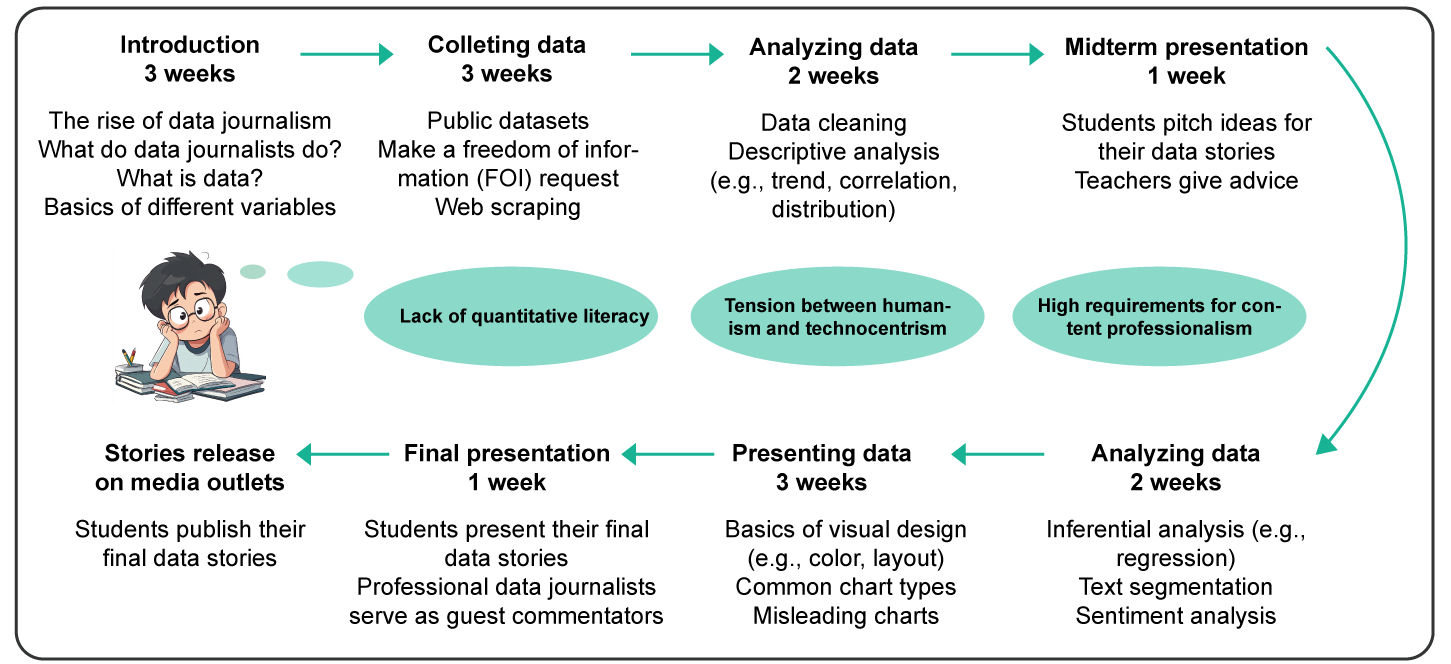}
 \centering
  \caption{An overview of the course structure presented in this work. The three challenges are discussed in Sec.~\ref{sec: char}.}
\label{fig:teaser}
}

\maketitle
\begin{abstract}
   The integration of data visualization in journalism has catalyzed the growth of data storytelling in recent years. Today, it is increasingly common for journalism schools to incorporate data visualization into their curricula. However, the approach to teaching data visualization in journalism schools can diverge significantly from that in computer science or design schools, influenced by the varied backgrounds of students and the distinct value systems inherent to these disciplines. This paper reviews my experience and reflections on teaching data-driven storytelling in a journalism school in Shanghai, China. To begin with, I discuss three prominent characteristics of journalism education (i.e., students' lack of quantitative literacy, the tension between humanism and technocentrism, and the high requirements for content professionalism) that pose challenges for course design and teaching. Then, for each challenge, I share firsthand teaching experiences and discuss corresponding approaches for teaching, such as trying to put visualization into a news context and finding commonality between data-driven storytelling and traditional storytelling. Overall, this paper aims to provide reference and inspiration for instructors who are teaching data visualization and data-driven storytelling to students with non-technical backgrounds.


\end{abstract}


\input{sections/01-intro}
\input{sections/02-context}

\input{sections/03-char}

\input{sections/04-teaching}

\input{sections/05-conclusion}

\bibliographystyle{eg-alpha-doi} 
\bibliography{egbibsample}       


\end{document}

%% file: sections/01-intro.tex
\section{Introduction}

Data visualization is now being used in teaching across various disciplines, with its application in journalism being notably representative. The fusion of these two fields is commonly referred to as data journalism, namely journalism done with data~\cite{bounegru2012data}, which requires ``a new set of skills for searching, understanding and visualizing digital sources in a time that basic skills from traditional journalism just aren’t enough''~\cite{bounegru2012data}. 
Since 2009, data journalism has entered a rapid development phase~\cite{history}. 
In Canada, Fred Vallance-Jones and David McKie published a book called \textit{Computer-Assisted Reporting: A Comprehensive Primer}~\cite{vallance2009computer}, emphasizing Canadian examples of computer-assisted reporting. In the UK, The Guardian established a dedicated data journalism column called Data Blog, which regularly updates data-driven stories in the form of information graphics~\cite{rogers2013facts}. Soon, data journalism flourished in news organizations worldwide, producing a series of innovative, interesting, and impactful data stories. This growth also spurred new research, such as Segel and Heer's seminal paper, \textit{Narrative visualization: Telling stories with data} (published in 2010)~\cite{segel2010narrative}. In this paper, researchers analyzed a corpus of data stories, most of which were published by news agencies, and proposed the first frameworks for data-driven visual narratives and narrative structures.

The rise of data journalism has root causes, with the profound impact of digitization on the media industry being a significant factor. Journalists are increasingly required to transcend traditional boundaries and embrace the challenges of the big data era~\cite{cohen2011computational,weber2018data}.
Unlike traditional journalism, data journalism demands a diverse skillset. While some newsrooms have successfully integrated data scientists or technicians with a passion for open data and hacker culture~\cite{usher2016interactive}, there remains a pressing need for journalism schools to educate students in adapting to this ``quantitative turn''~\cite{coddington2015clarifying} and prepare them as professional data journalists. In response, more and more data journalism courses and programs have been launched worldwide~\cite{dj_syllabuses,dj_courses}, such as the Data Journalism program (MS degree) at Columbia University~\cite{columbia} and the Journalism and Data Science program (BS degree) at Northeastern University~\cite{ne}. These programs generally incorporate interdisciplinary content, teaching students statistics, graphic design, as well as software and programming languages like R, Python, and JavaScript. Additionally, organizations such as The Centre for Investigative Journalism (based in the UK)~\cite{investigative_j} and summer bootcamps like The Lede Program~\cite{leed} are also committed to teaching journalists how to turn data into narratives. In this global trend, journalism schools in China have also begun to offer data journalism courses.

The reflections presented in this work are framed within this context. A key motivation of this paper is that teaching data visualization in journalism schools differs significantly from that in computer science or design schools, primarily due to the diverse backgrounds of students and the distinct characteristics and value systems of different disciplines. These differences profoundly influence curriculum design and pedagogy. Reflecting on these issues not only provides valuable evidence for the diverse applications of visualization in education but also illuminates how visualization, rooted mainly in computer science, can engage in dialogue with and empower social sciences.
In the following sections, I first introduce the background of the course I teach. Then, by combining my firsthand teaching experience and literature review, I discuss three prominent characteristics of teaching data visualization in journalism schools. For each characteristic, I share related teaching cases and recommend approaches for effective teaching.

%% file: sections/02-context.tex
\section{Teaching Context}

This paper shares the teaching experience of a course titled \textit{Data Analysis and Information Visualization} at the School of Journalism, Fudan University, Shanghai, China.
The School of Journalism at Fudan University is China's oldest institution for journalism education, established in 1929, renowned for its excellence in both teaching and research. Fudan University was also among the earliest in China to offer courses on data journalism, with \textit{Data Analysis and Information Visualization} being a prominent example. After years of development, this course has taught over one thousand students and has become a representative example of data journalism education in China. It has been recognized as a Nationally Distinguished Undergraduate Course by the Ministry of Education, which is the highest honor awarded to courses in China that serve as exemplary models nationwide.

Currently, this course is mandatory for all undergraduate students at the school, spanning 3 credits and consisting of 54 class hours. Approximately 200 students enroll each year, primarily from the journalism school, with a smaller number from other schools and departments such as literature and international politics.
Similar to the typical structure of data journalism courses~\cite{splendore2016educational}, this course covers topics including data collection, analysis, and visualization. The course utilizes a modular teaching approach, involving collaboration among three teachers. I am primarily responsible for teaching the data visualization module. At the beginning of the course, students need to form their own teams (normally 4-5 members per group). By the end of the course, each group should submit a data story (this final project accounts for about 60\% of the total grade), with outstanding works potentially being published in the news media. Due to the large number of students in the course, the class is primarily lecture-based. To encourage active participation, we have incorporated several interactive components: (i) classroom discussions. Students who actively engage in discussions will receive higher individual participation grades (about 10\% of the total grade). (ii) in-class individual quizzes. These quizzes test the knowledge points covered in class; after the quizzes, the teacher will explain the answers and address any questions (about 30\% of the total grade). (iii) group discussions. We allocate time after class to discuss with each group. Each group prepares several topics for proposal and engages in multiple rounds of discussions with the teachers to evaluate the value of their chosen topics. (iv) group presentations. At the end of the semester, each group is required to present their final work in class. We also invite industry experts to review the projects and provide feedback in person.

Throughout the teaching, we employ several methods to reflect on our teaching approaches and their effectiveness.
Firstly, we apply autoethnography methodology and regularly write memos to document our teaching experiences. These memos serve not only for self-reflection but also for exchange with teachers responsible for other modules during regular meetings, enhancing collaboration and improving the overall course quality.
Secondly, as the school requires all students to evaluate teachers and courses at the end of each semester, we receive approximately 200 ratings and textual feedback annually, providing valuable insights from the students' perspective.
Lastly, we keep abreast of developments in data journalism courses abroad and study relevant literature on data journalism education. Bhaskaran~\etal~\cite{bhaskaran2022teaching}, for instance, conducted a survey on data journalism education, identifying 29 papers published from 2014 to 2020. Building on this corpus, we employed the forward snowball method to include five additional papers published after 2020 (\ie ~\cite{bhargava2021teaching,georgiadou2023understanding,heravi2020data,kashyap2020teaching,kirchhoff2022journalism}). 
Interestingly, through surveying previous work, we often uncover common challenges faced by data journalism educators worldwide. Studying these international experiences provides valuable insights into how different educational institutions tackle similar issues. This enriches our understanding and informs our teaching practices, enabling us to implement proven strategies and adapt innovative solutions in our course curriculum.
In the following section, we will also integrate these papers to illustrate typical and generalizable characteristics  (rather than just isolated cases from a single school) of teaching data visualization at journalism schools.


%% file: sections/03-char.tex
\section{Characteristics of Journalism Education}
\label{sec: char}

Although we have an abundance of references nowadays for teaching data visualization, such as well-recognized textbooks~\cite{munzner2014visualization} and numerous books that teach how to create visualizations~\cite{rees2019survey}, it remains challenging to directly apply them in every context.
In the case of this work, a major challenge is that most of the aforementioned educational materials are written within the context of computer science. Consequently, their choice and organization of content typically follow engineering logic (\eg many common concepts in the visualization community, such as \textit{task}, \textit{pipeline}, \textit{input/output} are rooted in engineering discourse, which may not align with the thinking style of social sciences). This presents significant issues in determining what content to teach, which examples to use, and what research advancements should be introduced for journalism students.
On the other hand, journalism students possess their own knowledge structures and value systems. Teaching data visualization in journalism schools is embedded within the entire educational framework of journalism. Therefore, it is essential to first consider the characteristics of journalism education when designing the curriculum for data visualization courses.

By situating our first-hand teaching experience within the data storytelling model proposed by Lee~\etal\cite{lee2015more}, we identified and summarized three prominent challenges (marked as C1-C3 below) we encountered. These challenges correspond to the three main stages of data story creation: exploring data, making a story, and telling a story (see Fig. \ref{fig:teaser}).

\textbf{C1: Students' lack of quantitative literacy.}
A common phenomenon observed globally is that students in journalism schools often lack training in mathematics and computer science~\cite{dunwoody2013statistical,plaue2015data,treadwell2016numbers}. This gap in training restricts their ability and confidence to effectively handle and analyze data. As reviewed by Bhaskaran~\etal~\cite{bhaskaran2022teaching}, alleviating math anxiety in students is a common challenge faced by data journalism educators. Davies~\etal~\cite{davies2016data} interviewed journalism academics from 25 Australian universities about how they are incorporating data journalism into their courses, and 22 of the 25 respondents reported that math aversion among students was a problem.

In China, influenced by the high school education system, this problem is also pronounced. Since the 1970s, during high school, students have been segregated into arts or sciences, attending separate classes. In addition to the compulsory subjects of Chinese, math, and English, arts students usually learn subjects such as politics, history, and geography, while science students concentrate on physics, chemistry, and biology. In recent years, although this binary model of arts and sciences has begun to change, allowing students to choose subjects voluntarily in high school, those who choose journalism in college still tend to focus more on arts subjects.
In our classrooms, most students receive minimal quantitative training beyond high school math. Many students have developed a self-identity as ``not a math person'' and exhibit a fear of math and coding. This is evident in our class, where students frequently express concerns about handling data, both directly to teachers and on social media. Every semester, even after completing the course, students often use evaluation questionnaires to express complaints with quotes such as ``Why is this course so difficult?'', ``I'm studying journalism, so why do I have to learn coding?'', and ``I hate math'', reflecting their reluctance to engage with data.

\begin{figure}[t!]
 \centering
 \includegraphics[width=\columnwidth]{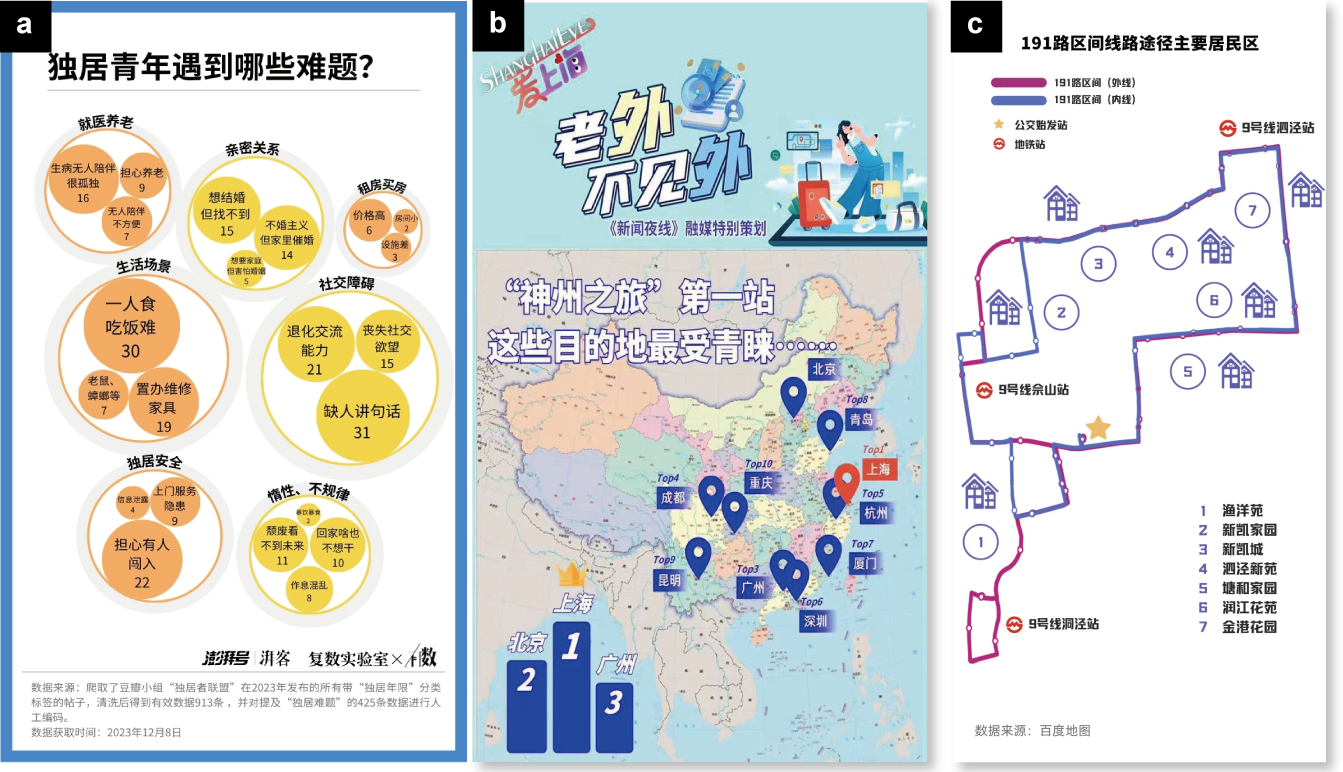}
 \caption{Examples of data stories created by students. (a) Title: ``Problems Encountered by Young People Living Alone.'' The bubble chart below summarizes the frequency of key topics on social media, such as social barriers and difficulties in daily life. (b) Title: ``Foreigners Are Not Outsiders.'' The map summarizes the most popular first destinations for foreigners traveling to China, with the top three being Shanghai, Beijing, and Guangzhou. (c) Title: ``Bus Passenger Flow by Time Period''. The students conducted on-site visits and recorded the passenger numbers at a Shanghai bus stop from 21:00 PM to 22:20 PM (represented by bars) as well as the number of people who were unable to board the last bus (represented by icons), reflecting on the inadequacy of late-night bus services for overtime workers.}
 \label{fig:students_work}
 \vspace{-2em}
\end{figure}


\textbf{C2: Tension between
humanism and technocentrism.}
The discipline of journalism traditionally places a strong emphasis on humanity rather than techniques, valuing skills such as writing, communication, filming, editing, and critical thinking. Compared to other humanities disciplines such as history and art, journalism often exhibits a stronger interest in social affairs. As society's watchdogs, journalists feel a strong sense of responsibility to oversee government, promote social justice, protect public interests, and defend minority rights~\cite{heravi20193ws,zion2014ethics}. For example, this can be clearly seen in the introduction of Columbia University's M.S. in Data Journalism, which states that the program ``provides the hands-on training needed to tell deeply reported data-driven stories \textit{in the public interest.}''~\cite{columbia}
In our class, topics such as social equality, public interest, and well-being are consistently among the most chosen subjects for data stories (see Fig. \ref{fig:students_work}). In other words, when making a story, journalism students do more than just arrange data insights logically. They also need to ensure that the whole story is compelling and holds significant meaning for society.

However, tension can arise between this humanistic tendency and the teaching of new technologies.
In fact, since the rise of data journalism, there have been constant concerns about technocentrism or even resistance from the journalism community (\eg ~\cite{parasie2015data,hewett2016learning}). Some fear that technical courses may push students toward becoming ``data workers'' who are glued to computers, potentially overshadowing journalism's core strengths, such as the ability to grasp and convey precise information, fact-checking, on-the-ground reporting, and investigative skills to uncover the truth. Some worry that journalism schools are losing their humanistic spirit and becoming boring technical training centers.
Therefore, there is a consensus that data journalism is ``not a replacement of traditional journalism, but an addition to it''~\cite{bounegru2012data}, and that we should strive for a balance between technical and humanistic aspects. 

\textbf{C3: High requirements for content professionalism.}
Journalism is a discipline focused on \textit{mass communication}, which involves large-scale communication typically aimed at the public. This contrasts with other scenarios, such as live presentations and individual or small-group presentations~\cite{kosara2013storytelling}.
Compared to these scenarios, mass communication is distinguished by its broad audience and potential for significant social impact when telling a story. Due to these traits, journalism has long emphasized professionalism~\cite{schudson2009objectivity}, requiring journalists to maintain a responsible attitude towards the public and carefully handle the content they create.
For example, the Society of Professional Journalists (SPJ, founded in 1909, is the oldest organization representing journalists in the United States) outlines ethical standards in its Code of Ethics~\cite{spj}, which include principles such as seeking truth and reporting it, minimizing harm, acting independently, and being accountable and transparent.

This emphasis extends to data journalism, where rigorous, transparent, and ethical data usage is highly valued~\cite{burns2018first,zion2014ethics}. For example, journalist Darrell Huff's seminal book \textit{How to Lie with Statistics}~\cite{huff2023lie}, published in the 1930s, exposes various deceptive visualization tactics found in newspapers.  This book vividly illustrates journalists' caution in using data to communicate with the public. Besides, consistent with journalism's tradition of verifying sources, data journalism places significant importance on attributing data. For example, when studying data stories, while the visualization community often focuses on design components (\eg ~\cite{segel2010narrative,stolper2018data}), journalism researchers would specifically examine data sources (\eg ~\cite{kashyap2020teaching,stalph2023exploring}).
Even if a visualization is beautifully crafted, its credibility is compromised if the data source is unclear, untrustworthy, or lacks authority.

%% file: sections/04-teaching.tex
\section{Teaching Approaches and Case Examples}
\label{sec:approach}

Next, corresponding to each of the above characteristics, I introduce specific teaching approaches and case examples (supplemental meterials such as example slides can be found at \url{https://osf.io/r76s8/}).

\subsection{For the lack of quantitative literacy (C1)}

\textbf{C1-1: Analogical thinking of data scales.}
Students' fear of data often originates from overly abstract numbers. To alleviate this fear, one method is to translate abstract scales into tangible forms, thus guiding students to engage with numbers in a more comfortable and enjoyable manner. 
Specifically, I often encourage students to engage in analogical thinking (\ie a way of understanding new or novel situations by building models based on existing knowledge of familiar scenarios~\cite{martin2003s}) by imagining data as something they feel relatable or by placing data in a physical context.

\begin{figure}[t!]
 \centering
 \includegraphics[width=0.9\columnwidth]{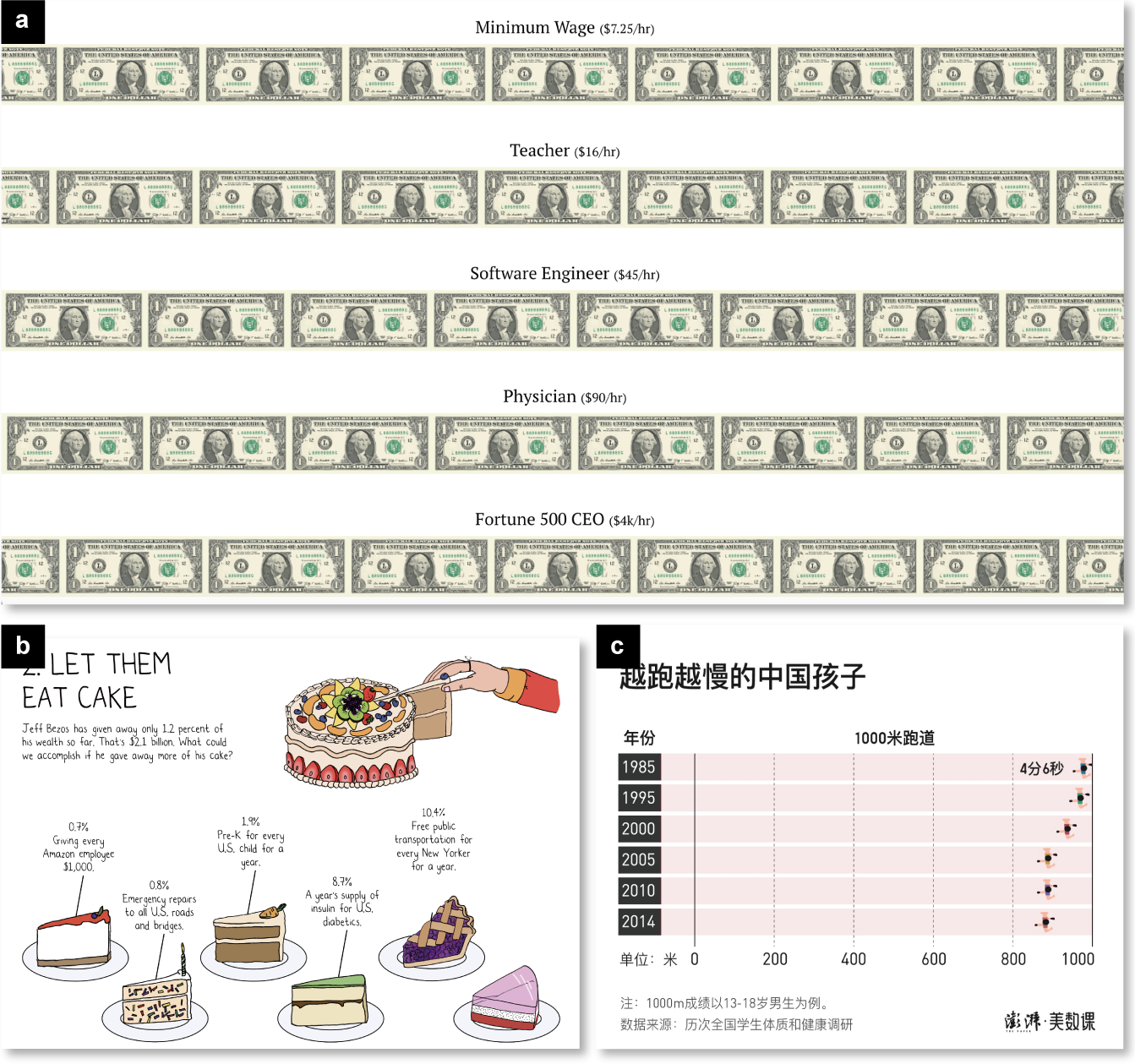}
 \caption{(a) Printing money; (b) 9 Ways to Imagine Jeff Bezos' Wealth; (c) Comparison of the time taken by Chinese boys aged 13-18 to complete a 1000-meter run from 1985 to 2014 (Title: ``The Slowing Pace of Chinese Children.'')}
 \label{fig:example_4}
 \vspace{-2em}
\end{figure}

For instance, financial reports often need to deal with money. When encountering large monetary scales like millions, billions, or trillions (such as reporting on billionaires' assets), people can easily be overwhelmed. So, how can we transform these scales into something understandable to the general public?
Given this problem, students are less likely to remain stuck in the fear of data, but will instead seek ways to solve it. Some students might think of analogies such as comparing a billionaire's hourly income to an average person's lifetime earnings, or estimating the number of trucks needed to transport his wealth. Often, the result is pleasantly surprising. Many of the ideas students come up with align remarkably well with some excellent visualization works. For instance, 
\textit{Printing money} (Fig. \ref{fig:example_4} (a)) uses a metaphor of a printing press to show the earning speed difference between the rich and average Americans, vividly portraying the wealth gap through animation. \textit{9 Ways to Imagine Jeff Bezos' Wealth} uses data comics to humorously depict Amazon CEO Bezos' wealth (Fig. \ref{fig:example_4} (b)). For example, with 0.7\% of his wealth, he could give each Amazon employee \$1000. With 0.8\% of his wealth, he could cover the emergency repair costs for all roads and bridges in the US. An Amazon employee's average annual income is \$37,930. To accumulate wealth equal to Bezos's, they would have to start working 4.5 million years ago. Such training also applies to China-specific topics, with the only difference being that for topics with distinct Chinese characteristics, it is necessary to consider scales familiar to the Chinese audience. For example, Fig. \ref{fig:example_4} (c) shows the decline in physical fitness among Chinese children over the past thirty years. To make this argument more tangible, the designer used a scale familiar to Chinese people, namely the time boys take to complete a 1,000-meter run. This is because all Chinese boys are required to undergo a 1,000-meter running test during their high school years. As shown in Fig. \ref{fig:example_4} (c), In 1985, Chinese boys averaged 4 minutes and 6 seconds for the 1,000-meter run, but by 2014, they could only cover 900 meters in the same time.
Through such training, students realize that data doesn't have to remain abstract; it can be interesting and engaging. They also learn that visualization extends beyond basic statistical charts and can be enjoyable to design.

\textbf{C1-2: Starting with spreadsheets and compatible tools.}
Similar to other educators~\cite{bradshaw2018data,treadwell2016numbers} , we have also been bothered by: which visualization tools should we teach? To what extent should we teach them so as not to overburden students? 
In fact, the challenge of teaching advanced tools has been observed in journalism classrooms globally~\cite{heravi20193ws}.
While professional tools such as Python and Tableau are popular in academia and excel at integrating various data science processes like data analysis and visualization, they present a steep learning curve for journalism students. Based on our observations, students strongly prefer Microsoft Excel for data analysis. As long as the data size is not large enough to cause Excel to crash, they tend to handle data in Excel first and then paste the prepared data into other tools. Directly selecting, manipulating, and transforming data in a spreadsheet helps students engage more directly with the data and manage its abstraction to some extent.
Therefore, teaching spreadsheet use for data processing, such as basic Excel operations, formulas for data cleaning, and pivot tables, has always been an important part of our course. We also emphasize the compatibility of visualization tools with spreadsheets to help students develop an effective workflow. For example, in text analysis of social media comments, students first organize the data in Excel, then use online tools like Wordclouds and Weiciyun (Chinese) for tokenization, frequency analysis, and creating word clouds. This approach better suits journalism students' skills and habits compared to coding in Python or R for the entire data scraping, analysis, and visualization process.

\subsection{For the pursuit of humanism (C2)}

\textbf{C2-1: Value-driven teaching strategy.}
Existing visualization textbooks typically follow a structure based on the technical pipeline of visualization, starting with techniques and then showcasing specific outputs (techniques -> results). The focus is on teaching \textit{how} to create data visualizations. However, for journalism students, the value of data visualization for people and society is more crucial. Before diving into techniques, students need to grasp \textit{why} learning them is essential. For educators, one effective approach is to reverse this above-mentioned order and begin with results before teaching techniques (results -> techniques). For instance, we can first present the social impacts that visualizations can achieve and how they contribute to public interest, thus fostering a deeper appreciation for the power of data visualization.
To achieve this goal, identifying compelling and relevant examples is crucial. Typically, I search for such examples from: (i) classic and well-known visualization cases that emphasize storytelling; (ii) contemporary cases with significant social impact, such as visualizations that have gone viral on social media or have been recognized by authoritative organizations; (iii) cases that have won prestigious awards in the journalism industry (\eg \textit{Snow Fall} by New York Times is the first interactive data story that won the Pulitzer Prize, and ``opened the door for new innovation in the newsroom''~\cite{snow}).

\begin{figure}[b!]
 \vspace{-1em}
 \centering
 \includegraphics[width=\columnwidth]{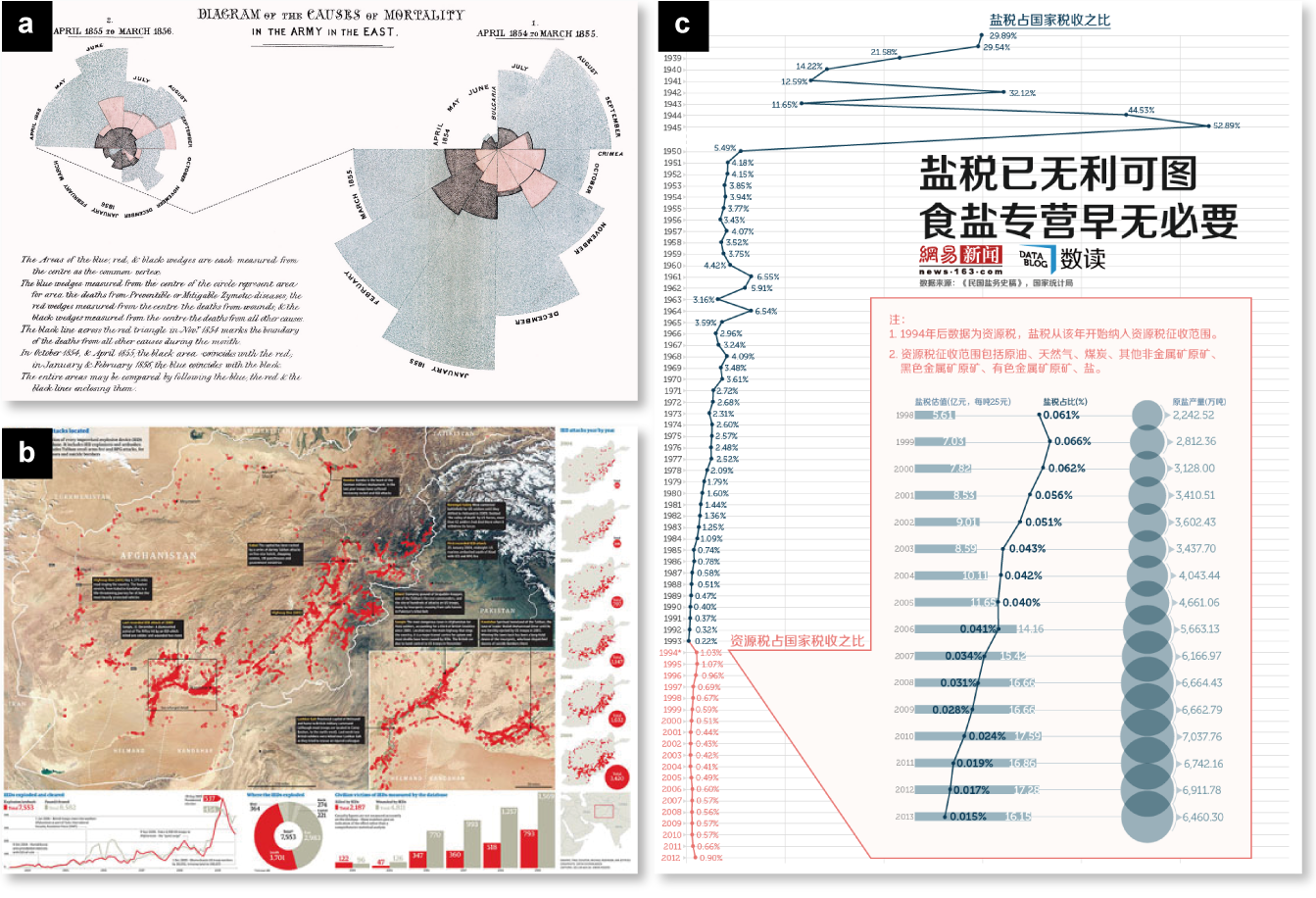}
 \caption{(a) Nightingale's rose chart; (b) Wikileaks Afghanistan files: every IED attack, with co-ordinates; (c) The line chart reflects the proportion of the salt tax in China's total tax revenue, which keeps going done. (Title: ``The salt tax is no longer profitable, and the monopoly on salt is no longer necessary.'' )}
 \label{fig:example_1}
\end{figure}

For example, regarding (i), Florence Nightingale's rose chart is a vivid example. During the Crimean War (1853-1856), Nightingale observed that many soldiers did not die directly from battlefield injuries but succumbed due to poor medical conditions after being transported to rear camps.
To bring this issue to the attention of British officials and prompt action to improve medical care, Nightingale meticulously created a rose chart. This innovative visualization, featured in a report submitted to the government, displayed the number of soldier deaths attributed to various causes in a stacked fan format arranged chronologically.
The rose chart is not only visually striking and novel but also effectively highlights differences in data, particularly emphasizing the outer circle representing soldiers who perished due to poor medical conditions (Fig. \ref{fig:example_1} (a)). Nightingale successfully achieved her objective through this visualization, influencing policymakers of her time and catalyzing societal change. This narrative captivates journalism students, illustrating the influential power of storytelling through data visualization and its relevance to their professional pursuits.

Another example from contemporary times is WikiLeaks. In 2010, this non-profit organization released over 250,000 US diplomatic cables. Investigative journalist Stefania Maurizi described it as, ``for the first time in history, WikiLeaks has ripped a gaping hole in this secret power, giving billions of people systematic and unrestricted access to enormous archives of classified documents revealing how our governments behave when, completely shielded from public and media scrutiny, they prepare wars or commit atrocities.''~\cite{wiki}
Based on this massive dataset, journalists mined insights and presented them to the public through a series of influential data stories such as The Guardian's graphical reports~\cite{wiki_guardian} (Fig. \ref{fig:example_1} (b)) and The New York Times' war logs~\cite{wiki_nyt}. For instance, some stories visualized top-secret war attack records on maps, some depicted relationships among vast documents using node-link diagrams, and some provided cleaned datasets for public download. Learning about this case often ignites students' enthusiasm to pursue journalistic missions through big data. They quickly recognize the value of visualization in making complex datasets readable, understandable, and capable of revealing stories that traditional reporting methods would struggle to handle.

We also use cases from the Chinese context to inspire students' enthusiasm for addressing domestic issues that are significant to the Chinese people. For example, in 2016, China abolished the salt monopoly policy, ending the Chinese government's over two-thousand-year control over the production and distribution of salt. In ancient China, salt was always one of the most profitable commodities, and this policy provided substantial revenue to the government, supporting the vast expenditures of the country. So, why did we decide to abolish this system now? Would it disrupt market prices or impact household living costs? These were the concerns of billions of Chinese people. To address these questions, Chinese media used data visualization to provide answers. As shown in Fig. \ref{fig:example_1} (c), this vertical line chart illustrates how the proportion of salt tax in China’s total tax revenue has changed over time. It clearly shows that the importance of the salt tax has steadily declined, and in recent years, its contribution to national revenue has become negligible. The salt monopoly had thus lost its economic necessity. Through this concise infographic, the reasons behind the economic policy were effectively communicated to the public.

\begin{figure}[b!]
\vspace{-2em}
 \centering
 \includegraphics[width=\columnwidth]{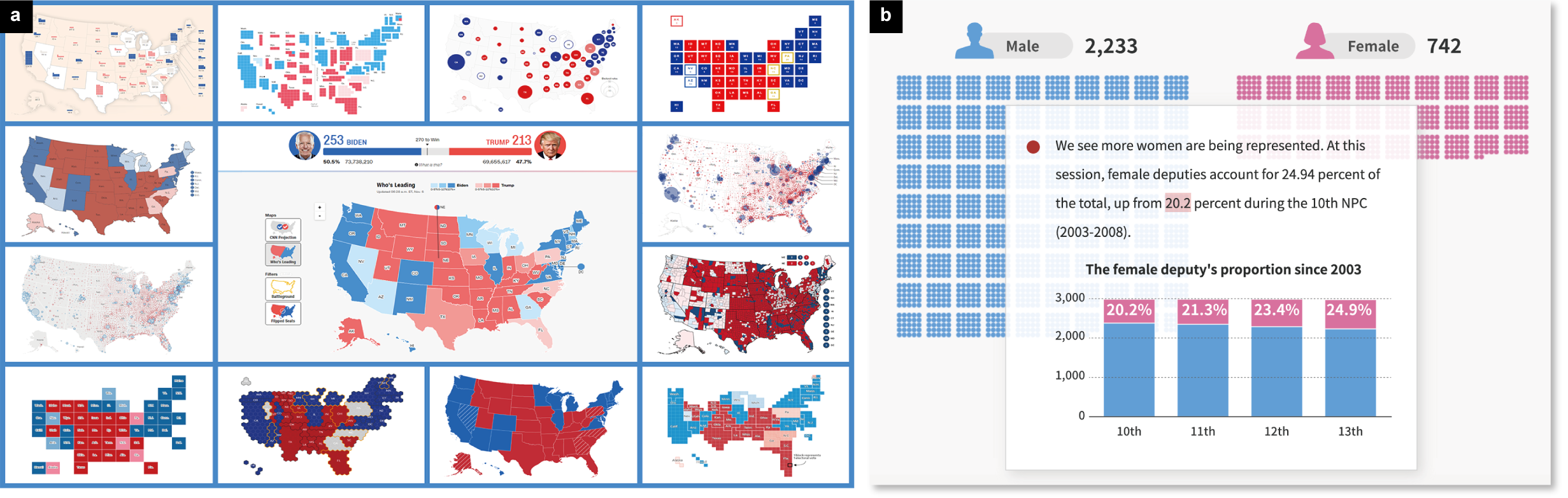}
 \caption{(a) Various visualizations of US presidential election results; (b) Visualizations of China's NPC representatives.}
 \label{fig:example_2}
 \vspace{0em}
\end{figure}

\textbf{C2-2: Putting visualization into a news context.}
The ultimate goal of teaching data visualization in journalism schools is to create compelling data stories. For any story, context is indispensable. Therefore, in our class, instead of directly teaching students visualization design principles and methods, we employ various storytelling scenarios to guide their learning of visualization design. This approach echoes what Graham referred to as the ``guerrilla approach''~\cite{graham2018diy}.
For example, we often begin with trending news topics to stimulate brainstorming sessions. Students are presented with specific reporting scenarios, such as, ``Imagine you're covering the U.S. presidential election and have voting data from various states. How would you visualize this information?'' Initially, students might consider using a choropleth map. However, we then challenge them to think critically: ``Is the choropleth map ideal? Are there potential pitfalls?'' Some students may realize that states with large geographical areas but sparse populations could appear more influential than they actually are, potentially misleading in portraying election results.
Identifying such issues naturally leads to considering alternative visualization choices, allowing teachers to introduce concepts like cartograms, which prioritize conveying quantitative information over exact geographical accuracy, along with their specific design methods~\cite{nusrat2016evaluating}.
Fig. \ref{fig:example_2} (a) illustrates various design options discussed in our class, such as simplifying states as equally-sized squares and overlaying bars on the map. I guide students through detailed analyses of these design alternatives, encouraging them to understand the rationale behind each design and weigh their advantages and disadvantages.

Given the significant differences between Chinese society and Western countries, we also guide students to explore news reporting in China. For example, unlike the U.S. election system mentioned earlier, China’s election system features multiple tiers of elected representatives, culminating in the National People's Congress (NPC), which meets annually to set national policies and laws.
However, Chinese media previously only published text lists of these representatives, lacking visual enhancements. Therefore, similar to the earlier case with the U.S. election, I ask students to imagine: ``If you were covering the NPC meeting, how would you present information about these representatives? How can you move beyond simple text or tables?'' Based on this setting, we discuss and analyze various feasible visual design options. For example, statistical charts could help show the representatives' identities. Or, each representative could be depicted as a unit, and users could interact by hovering to view more detailed information (Fig. \ref{fig:example_2} (b)).
Such context-driven training enables students to develop a deeper understanding of effective visualizations. They also gain proficiency in utilizing different visual channels and optimizing data encodings to enhance storytelling.


\textbf{C2-3: Finding commonality with traditional storytelling.}
Data storytelling lies at the intersection of data science and traditional storytelling (\eg narratology, cinematology). As mentioned earlier, there is both tension (\eg different thinking styles and terminologies) and complementary attraction between these fields. For instance, journalism students usually excel at creating storylines and crafting compelling stories. However, they often lack proficiency in handling data. In contrast, the field of visualization excels at handling data and structurally breaking down data stories as, for example, a series of design spaces driven by specific tasks (\eg ~\cite{yang2021design,lan2021understanding,shi2021communicating}). Such research yields concrete guidelines for data storytellers and often provides valuable teaching materials (\eg online taxonomies and galleries), but it is relatively uncommon in journalism education.
Therefore, an essential teaching approach is to impart visualization knowledge while considering the existing knowledge structure of journalism students.

\begin{figure}[t!]
 \centering
 \includegraphics[width=0.78\columnwidth]{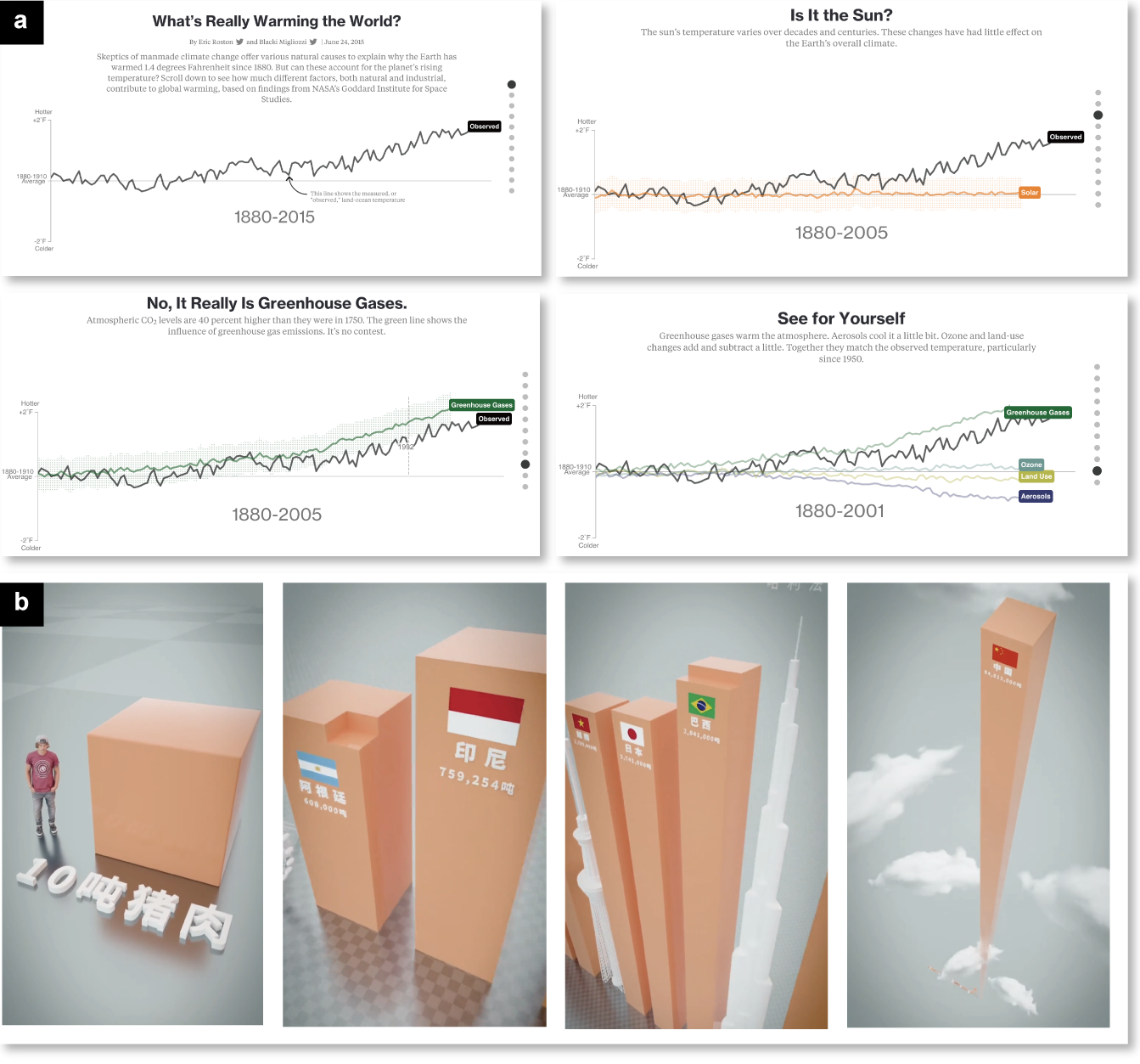}
 \caption{(a) What's really warming the world?; (b) Do you know how much pork we eat each year? (the Chinese phrase in the first picture means ``10 tons of pork'')}
 \label{fig:example_3}
 \vspace{-3em}
\end{figure}

To meet this goal, we prioritize introducing research that intersect with traditional narrative studies in our class. For example, Lan~\etal~\cite{lan2021understanding} incorporated temporal concepts such as chronology, flash-forwards, and flashbacks in data story research. Shi~\etal~\cite{shi2021communicating} summarized animation techniques in data stories and categorized them into different narrative strategies such as creating tension, twists, and maintaining coherence.
Yang~\etal~\cite{yang2021design} summarized how pyramid narrative structures can be applied in data storytelling. An exemplary case from this work is Bloomberg's \textit{What's really warming the world?} (Fig. \ref{fig:example_3} (a)).
It starts with a provocative question about the causes of global warming, creating an intriguing hook. The story then sequentially presents correlations between different variables (\eg sun, volcanoes) and temperature, building tension. It culminates in revealing the true cause: greenhouse gas emissions, reaching the climax of the narrative. Finally, the story summarizes all the data, providing resolution.
Another case more relevant in the Chinese context is a data video about pork consumption (Fig. \ref{fig:example_3} (b)). The video begins with countries that have the least pork consumption. As the camera pulls up, the pork consumption of additional countries is presented one by one, gradually building emotional impact. Next, the camera rapidly zooms out to reveal China's towering bar and enormous pork consumption, reaching the narrative climax.
These cases engage students, allowing them to remain connected to familiar narrative theories while also receiving guidance on how to construct compelling narratives for data.

We also encourage students to combine traditional reporting methods, such as interviewing and field investigations, with data gathering. For example, a group of students conducted a field investigation on the insufficient capacity of late-night buses in Shanghai (Fig. \ref{fig:students_work}(c)). The students took to the streets, observing and recording data on nighttime passenger flow and bus schedules at various stations. They effectively balanced the quantitative methods of data journalism with the traditional journalism focus on field reporting. Their data story garnered attention from both the public and the government, successfully leading to an increase in bus services. Another group of students created a data story themed around the inconveniences foreigners face when traveling in China. They not only analyzed datasets on the number of foreign tourists to China and flight information but also audited the support for English interfaces and foreign currency payments across various commonly used Chinese apps. Additionally, they conducted street interviews with foreign tourists in Shanghai, effectively combining quantitative and qualitative methods.


\subsection{For the requirements of content professionalism (C3)}

\begin{figure}[t!]
 \centering
 \includegraphics[width=\columnwidth]{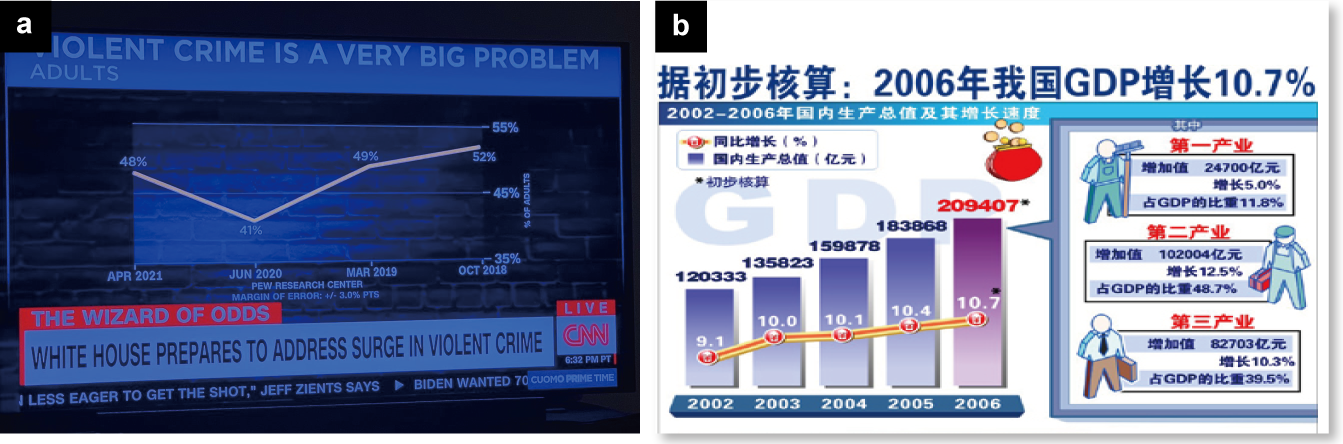}
 \caption{(a) A reversed x-axis; (b) Disproportionate bars and the missing of axes (The title means ``According to preliminary calculations, China's GDP grew by 10.7\% in 2006'').}
 \label{fig:example_5}
 \vspace{-2em}
\end{figure}

\textbf{C3-1: Creating visualizations that are trustworthy and suitable for public understanding.}
Due to journalism's public service nature, we dedicate considerable class time to discussing design flaws in visualizations and the harms they cause. This includes commonly discussed perceptual flaws (\eg truncated axes, inverted axes, 3D distortions, misuse of colors), as well as reasoning flaws such as cherry-picking data and normalization errors~\cite{lan2024came}. When teaching these concepts, we particularly focus on real-world examples gathered from news media, such as problematic visualizations in newspapers and TV shows (\eg Fig. \ref{fig:example_5}). Potential sources of such cases include books written by data journalists~\cite{cairo2019charts,wong2013wall}, corpus contributed by previous research~\cite{lo2022misinformed,lan2024came}, and specialized events such as VisLies~\cite{vislie}.) Such cases can also serve as classroom quizzes, requiring students to detect flaws and suggest improvements. This training effectively cultivates students' critical abilities in visualization design and enhances their sensitivity to identifying design flaws in real-life situations.

Additionally, to create visualizations that are suitable for public communication, we recommend visual vocabularies such as those summarized by the Financial Times~\cite{ft} to help students explore visualization types commonly used by journalists. We also dedicate substantial time (about 10 class hours) to teaching tools like \textbf{Flourish, Datawrapper, Canva, and Dycharts (Chinese)} to implement these visualizations. These tools are mostly template-based. In addition to being easy to learn, we see two other reasons why they are particularly suitable for journalism students.
Firstly, while tools like D3.js and Tableau support flexible visual mappings and enable highly creative designs, this level of freedom may exceed the needs of journalists. Visualizations created by journalists are primarily aimed at the general public and must consider their literacy. Consequently, visualizations in news media tend to be conventional and not overly complex, resulting in relatively low motivation among journalism students to create highly customized charts.
Secondly, tools like Flourish are evolving to meet the needs of journalists. In addition to providing a wide range of chart types, they now offer advanced templates such as animated bar races, population pyramids, and specialized visualizations for reporting sports events like swimming and running. These tools also support various export formats (\eg static images, HTML embeddings), enabling journalists to meet most of their visualization needs.


\textbf{C3-2: Designing for content dissemination.}
Effective mass communication requires careful planning of the media and channels for content dissemination.
For instance, media types may encompass print media (\eg newspapers, magazines), digital media (\eg websites, blogs), broadcast media (\eg television, radio), outdoor media (\eg billboards, signage), and other emerging media (\eg AR/VR), each leading to varying story genres and publication features~\cite{lan2023affective,segel2010narrative}. 
Within a certain media type, the characteristics of specific devices should also be taken into account. For instance, digital media often demands adaptability across various devices, such as PCs and mobile phones. This necessitates adjustments like optimizing aspect ratios for mobile screens and addressing interaction challenges, such as the fat finger issue~\cite{kim2021design}. Students are thus encouraged to consider questions such as: Which media type should I use? Where should I publish my story? Should I adopt cross-media strategies? Does my visualization design fulfill these objectives? We encourage students to utilize mind maps to navigate these questions and evaluate whether their stories effectively address them. For example, Fig. \ref{fig:students_work} (f) was originally published on WeChat as a long article composed of text, data visualizations, and a street interview video. To enhance content dissemination, they published the video individually on short video apps and repurposed the main content of the article into an infographic. Their work was later noticed by Shanghai Television, which invited the students to participate in a live TV broadcast to present their visualizations and findings. This demonstrates that, under the different distribution needs of news content, the design of data stories can be multi-version and multi-genre.




\textbf{C3-3: Familiarizing with publication policies and auditing pitfalls/risks.}
To ensure students' work adheres to publication standards and mitigates potential risks, we invite industry mentors (\eg heads of data journalism teams) to review and rate students' stories during the final class. Apart from assessing whether students' writing and graphics comply with industry standards, mentors leverage their extensive industry experience to evaluate the legal risks associated with publication. This is especially important in the Chinese context. For example, in China, map publications must follow the National Map Management Regulations. 
Maps that are inconsistent with China's political and diplomatic positions (\eg incorrect border lines, mislabeling certain provinces or regions as countries, or omitting islands) can lead to severe consequences, such as publication retractions, public controversy, or even legal action. Thus, mentors will remind students of the problems with their maps and suggest that they download map templates and GIS files from official websites.
Additionally, mentors provide insights into platform-specific regulations and best practices. For instance, commercial platforms like TikTok and Douyin (Chinese) have unique algorithmic guidelines for content recommendations. By leveraging this knowledge, mentors help students optimize their titles, charts, and keywords to ensure compliance with platform rules, maximize visibility, and minimize the risk of content removal or account bans. For official mainstream media, adhering to government policies is crucial. Data stories that align with national ideologies and major policies (\eg protecting cultural heritage, enhancing cultural confidence, and promoting AI development) are more likely to be favored and recommended by editors.

%% file: sections/05-conclusion.tex
\section{Discussion and Conclusion}

Echoing previous literature~\cite{cairo2012functional,hewett2016learning,splendore2016educational}, data journalists have their own characteristics and preferences when using visualizations. For example, compared to engineers, they pay more attention to the narrative significance and aesthetics, believing that form should serve the content. They do not particularly pursue technical complexity, but rather hope that visualizations can truly play an effective role in communicating with the public. This demonstrates that, as the application of visualization becomes increasingly diverse, the teaching of visualization also needs to be adapted in a timely manner to meet the needs of various fields and disciplines.

Although this paper primarily reflects on a specific course at a university in Shanghai, some of the findings may have generalizable significance. Apart from journalism, the shift to data is a global phenomenon, and many disciplines are currently incorporating data-related courses. For instance, digital humanities has led historians, linguists, literary scholars, and artists to engage in data processing and mining. In science and engineering, big data and visualization are widely used in fields like medicine and biology. Students in these fields, who may lack a background in data visualization, can also face steep learning curves or paradigm conflicts. Thus, the methods discussed in this work, such as integrating visualization teaching with students' existing knowledge structures and starting with tools that students find easy to onboard, can in fact be mapped to the more general ideas of social constructivism~\cite{amineh2015review}. This theory, which introduced the concept of the Zone of Proximal Development, suggests that teachers should play the role of scaffolding, helping students to move from their current level to a new level through careful guidance and appropriate pedagogical design.

But of course, this developmental process can be challenging. In our course, for example, significant variation among students often results in polarized feedback in end-of-course surveys. While some students report gaining valuable technical skills, others still find the course too difficult. Although our introduced teaching strategies have generally improved classroom participation and post-course evaluations, individual differences cannot be mitigated. Therefore, apart from paying greater attention to students' diverse learning states and providing timely assistance, we have also carefully designed students' grading criteria, placing greater emphasis on the final project score. For example, a student strong in storytelling but weaker in data analysis quizzes can still contribute significantly to the team and earn a good grade. This approach aligns with the data journalism industry, where newsrooms are composed of individuals with diverse strengths, and proficiency in every aspect is usually not required~\cite{rogers2013facts,wong2013wall}. We believe that encouraging everyone to leverage their strengths rather than penalizing their weaknesses should be the true essence of such interdisciplinary courses.

There are still many issues to be discussed and researched. For instance, the geographical shift in data journalism research from Western countries to the Global South~\cite{mutsvairo2019challenges,wright2023development} highlights the need to investigate unique challenges posed by different cultures, audiences, social media platforms, and government intervention in these regions. Although this work has integrated some China-specific teaching experiences and examples, more issues await future exploration. Besides, the rise of AI is challenging the established teaching models of data journalism~\cite{heravi20193ws,splendore2016educational}. The power of AI in data analysis and visualization may potentially benefit students who find coding and statistics daunting. However, this technological advancement also necessitates that teachers restructure their syllabi and the tools they teach. Currently, our teaching team is exploring ways to enhance traditional teaching methods through human-AI collaboration. We also anticipate more research exploring AI-driven approaches to visualization education.

\section*{Acknowledgement}
This work was supported in part by the National Natural Science Foundation of China 62402121, Shanghai Chenguang Program, and Research and Innovation Projects from the School of Journalism at Fudan University. 

\appendix
\section{Image Credits}

Fig. \ref{fig:teaser}: Adapted from the image by Lee~\etal~\cite{lee2015more}.

\noindent Fig. \ref{fig:students_work} (a): Yanni Li, Shuyi Li, Lei Lei, Xiuqi Tian, and Xiaochun Wang (Fudan University), from \url{https://www.thepaper.cn/newsDetail_forward_28189342}.

\noindent Fig. \ref{fig:students_work} (b): Ruoxin Cao, Haoning Ma, Bo Pang, Jiahui Wu, and Ningyu Yang (Fudan University), from \url{https://www.thepaper.cn/newsDetail_forward_26912005}.

\noindent Fig. \ref{fig:students_work} (c): Yu Wang, Zekai Tang, Jiajun Xiang, Jianghua Zhang, and Yuan Yao (Fudan University), from 
\url{https://www.shobserver.com/staticsg/res/html/web/newsDetail.html?id=720450}.

\noindent Fig. \ref{fig:students_work} (d): Qiaoli Ding, Yi Dai, Niyi Gu, Zhouyi Yao, and Yixin Zhang (Fudan University), from \url{https://www.thepaper.cn/newsDetail_forward_26439072}.

\noindent Fig. \ref{fig:students_work} (e): Mengxin Xie, Yaqi Wei, Weiqing Wang, Jiaqi Zhu, and Yuyao Zuo (Fudan University), from \url{https://m.thepaper.cn/newsDetail_forward_27815807}.

\noindent Fig. \ref{fig:students_work} (f): Yinuo Chen, Jiameng Guo, Yujia Peng, Yuxin Wang, and Zhou Yang (Fudan University), from \url{https://www.kankanews.com/detail/dZ2e8N0ZpwR}.

\noindent Fig. \ref{fig:example_4} (a): Neal Agarwal, from \url{https://neal.fun/printing-money/}.

\noindent Fig. \ref{fig:example_4} (b): Mona Chalabi, from \url{https://www.nytimes.com/interactive/2022/04/07/magazine/jeff-bezos-net-worth.html}.

\noindent Fig. \ref{fig:example_4} (c): Xiner Jiang, Yao Wei, Yasai Wang, and Shujing Zheng (thePaper), from \url{https://www.thepaper.cn/newsDetail_forward_14547210}.

\noindent Fig. \ref{fig:example_1} (a): Florence Nightingale, from \url{https://en.wikipedia.org/wiki/File:Nightingale-mortality.jpg}.

\noindent Fig. \ref{fig:example_1} (b): Simon Rogers (The Guardian), from \url{https://www.theguardian.com/news/datablog/2011/jan/31/wikileaks-data-journalism}.

\noindent Fig. \ref{fig:example_1} (c): Yabin Zhang, Xiaoqi Wu (NetEase), from \url{https://web.archive.org/web/20171115004130/http://data.163.com/14/1121/08/ABIFVLMM00014MTN.html}.

\noindent Fig. \ref{fig:example_2} (a): Curated and designed by AnyChart, from \url{https://www.anychart.com/blog/2020/11/06/election-maps-us-vote-live-results/}.

\noindent Fig. \ref{fig:example_2} (b): Jing Li, Daqiao Lin, Ling Lin, Yangjijing Liu, Shuhui Peng, and Xinrui Wang (Nanjing University), from \url{https://mp.weixin.qq.com/s/s5HIxhauiHmXQfRiCkNFjg}.

\noindent Fig. \ref{fig:example_2} (c): CGTN, from \url{https://news.cgtn.com/event/2019/whorunschina/index.html/}.

\noindent Fig. \ref{fig:example_3} (a): Eric Roston and Blacki Migliozzi (Bloomberg), from \url{https://www.bloomberg.com/graphics/2015-whats-warming-the-world/}.

\noindent Fig. \ref{fig:example_3} (b): Jin10 Data, from \url{https://v.jin10.com/details.html?id=10749}.

\noindent Fig. \ref{fig:example_5} (a): an anonymous contributor to WTFViz, from \url{https://viz.wtf/post/658882425666011136/the-x-axis-is-supposed-to-be-backwards-right}.

\noindent Fig. \ref{fig:example_5} (b): Xinhua, from \url{https://www.gov.cn/2007lh/content_537557.htm}.

%% file: main.bbl
\newcommand{\etalchar}[1]{$^{#1}$}
\begin{thebibliography}{\uppercase{SDSE{\etalchar{*}}16}}

\bibitem[AA15]{amineh2015review}
\textsc{Amineh R.~J., Asl H.~D.}:
\newblock Review of constructivism and social constructivism.
\newblock \emph{Journal of social sciences, literature and languages 1}, 1 (2015), 9--16.

\bibitem[Bah22]{snow}
\textsc{Bahr S.}:
\newblock ‘snow fall’ at 10: How it changed journalism.
\newblock \url{https://www.nytimes.com/2022/12/23/insider/snow-fall-at-10-how-it-changed-journalism.html}, 2022.
\newblock Last Access: 2025-01-05.

\bibitem[BCG12]{bounegru2012data}
\textsc{Bounegru L., Chambers L., Gray J.}:
\newblock \emph{The data journalism handbook}.
\newblock European Journalism Centre, 2012.

\bibitem[BD21]{bhargava2021teaching}
\textsc{Bhargava R., D'Ignazio C.}:
\newblock Teaching data journalism in a world of tool and tech overload.
\newblock In \emph{Companion Publication of the ACM Web Science Conference} (2021), pp.~17--22.
\newblock \href {https://doi.org/10.1145/3462741.3466660} {\path{doi:10.1145/3462741.3466660}}.

\bibitem[BKM22]{bhaskaran2022teaching}
\textsc{Bhaskaran H., Kashyap G., Mishra H.}:
\newblock Teaching data journalism: A systematic review.
\newblock \emph{Journalism Practice 18}, 3 (2022), 722--743.
\newblock \href {https://doi.org/10.1080/17512786.2022.2044888} {\path{doi:10.1080/17512786.2022.2044888}}.

\bibitem[BM18]{burns2018first}
\textsc{Burns L.~S., Matthews B.~J.}:
\newblock First things first: Teaching data journalism as a core skill.
\newblock \emph{Asia Pacific Media Educator 28}, 1 (2018), 91--105.
\newblock \href {https://doi.org/10.1177/1326365X18765530} {\path{doi:10.1177/1326365X18765530}}.

\bibitem[Bra18]{bradshaw2018data}
\textsc{Bradshaw P.}:
\newblock Data journalism teaching, fast and slow.
\newblock \emph{Asia Pacific Media Educator 28}, 1 (2018), 55--66.
\newblock \href {https://doi.org/10.1177/1326365X18769395} {\path{doi:10.1177/1326365X18769395}}.

\bibitem[Cai12]{cairo2012functional}
\textsc{Cairo A.}:
\newblock \emph{The Functional Art: An introduction to information graphics and visualization}.
\newblock New Riders, 2012.

\bibitem[Cai19]{cairo2019charts}
\textsc{Cairo A.}:
\newblock \emph{How charts lie: Getting smarter about visual information}.
\newblock WW Norton \& Company, 2019.

\bibitem[CHT11]{cohen2011computational}
\textsc{Cohen S., Hamilton J.~T., Turner F.}:
\newblock Computational journalism.
\newblock \emph{Communications of the ACM 54}, 10 (2011), 66--71.
\newblock \href {https://doi.org/10.1145/2001269.2001288} {\path{doi:10.1145/2001269.2001288}}.

\bibitem[Cod15]{coddington2015clarifying}
\textsc{Coddington M.}:
\newblock Clarifying journalism’s quantitative turn: A typology for evaluating data journalism, computational journalism, and computer-assisted reporting.
\newblock \emph{Digital Journalism 3}, 3 (2015), 331--348.
\newblock \href {https://doi.org/10.1080/21670811.2014.976400} {\path{doi:10.1080/21670811.2014.976400}}.

\bibitem[{Col}a]{leed}
\textsc{{Columbia University}}:
\newblock The lede program.
\newblock \url{https://ledeprogram.com/}.
\newblock Last Access: 2025-04-01.

\bibitem[{Col}b]{columbia}
\textsc{{Columbia University}}:
\newblock {M.S. Data Journalism}.
\newblock \url{https://journalism.columbia.edu/ms-data-journalism}.
\newblock Last Access: 2025-01-05.

\bibitem[DC16]{davies2016data}
\textsc{Davies K., Cullen T.}:
\newblock Data journalism classes in australian universities: Educators describe progress to date.
\newblock \emph{Asia Pacific Media Educator 26}, 2 (2016), 132--147.
\newblock \href {https://doi.org/10.1177/1326365X16668969} {\path{doi:10.1177/1326365X16668969}}.

\bibitem[DG13]{dunwoody2013statistical}
\textsc{Dunwoody S., Griffin R.~J.}:
\newblock Statistical reasoning in journalism education.
\newblock \emph{Science Communication 35}, 4 (2013), 528--538.
\newblock \href {https://doi.org/10.1177/1075547012475227} {\path{doi:10.1177/1075547012475227}}.

\bibitem[{Fin}]{ft}
\textsc{{Finantial Times}}:
\newblock Visual vocabulary.
\newblock \url{https://ft-interactive.github.io/visual-vocabulary/}.
\newblock Last Access: 2025-01-05.

\bibitem[GM23]{georgiadou2023understanding}
\textsc{Georgiadou E., Matsiola M.}:
\newblock Understanding and enhancing journalism students’ perception of data journalism.
\newblock \emph{Journalism and Media 4}, 4 (2023), 1232--1247.
\newblock \href {https://doi.org/10.3390/journalmedia4040078} {\path{doi:10.3390/journalmedia4040078}}.

\bibitem[Gra18]{graham2018diy}
\textsc{Graham C.}:
\newblock A diy, project-based approach to teaching data journalism.
\newblock \emph{Asia Pacific Media Educator 28}, 1 (2018), 67--77.
\newblock \href {https://doi.org/10.1177/1326365X18768308} {\path{doi:10.1177/1326365X18768308}}.

\bibitem[Her19]{heravi20193ws}
\textsc{Heravi B.~R.}:
\newblock 3ws of data journalism education: What, where and who?
\newblock \emph{Journalism Practice 13}, 3 (2019), 349--366.
\newblock \href {https://doi.org/10.1080/17512786.2018.1463167} {\path{doi:10.1080/17512786.2018.1463167}}.

\bibitem[Hew16]{hewett2016learning}
\textsc{Hewett J.}:
\newblock Learning to teach data journalism: Innovation, influence and constraints.
\newblock \emph{Journalism 17}, 1 (2016), 119--137.
\newblock \href {https://doi.org/10.1177/1464884915612681} {\path{doi:10.1177/1464884915612681}}.

\bibitem[HL20]{heravi2020data}
\textsc{Heravi B.~R., Lorenz M.}:
\newblock Data journalism practices globally: Skills, education, opportunities, and values.
\newblock \emph{Journalism and Media 1}, 1 (2020), 26--40.
\newblock \href {https://doi.org/10.3390/journalmedia1010003} {\path{doi:10.3390/journalmedia1010003}}.

\bibitem[Hou21]{history}
\textsc{Houston B.}:
\newblock The history of data journalism.
\newblock \url{https://datajournalism.com/read/longreads/the-history-of-data-journalism}, 2021.
\newblock Last Access: 2025-01-05.

\bibitem[Huf23]{huff2023lie}
\textsc{Huff D.}:
\newblock \emph{How to lie with statistics}.
\newblock Penguin UK, 2023.

\bibitem[Jam22]{wiki}
\textsc{Jamal E.~O.}:
\newblock How wikileaks revolutionised the world of journalism.
\newblock \url{ https://www.thedailystar.net/opinion/views/the-overton-window/news/how-wikileaks-revolutionised-the-world-journalism-3193351}, 2022.
\newblock Last Access: 2025-01-05.

\bibitem[jwy]{dj_courses}
\textsc{jwyg}:
\newblock Data journalism university courses and programmes.
\newblock \url{https://github.com/jwyg/data-journalism-courses/blob/master/data-journalism-university-courses.csv}.
\newblock Last Access: 2025-04-01.

\bibitem[KB20]{kashyap2020teaching}
\textsc{Kashyap G., Bhaskaran H.}:
\newblock Teaching data journalism: insights for indian journalism education.
\newblock \emph{Asia Pacific Media Educator 30}, 1 (2020), 44--58.
\newblock \href {https://doi.org/10.1177/1326365X20923200} {\path{doi:10.1177/1326365X20923200}}.

\bibitem[Kir22]{kirchhoff2022journalism}
\textsc{Kirchhoff S.}:
\newblock Journalism education’s response to the challenges of digital transformation: A dispositive analysis of journalism training and education programs.
\newblock \emph{Journalism Studies 23}, 1 (2022), 108--130.
\newblock \href {https://doi.org/10.1080/1461670X.2021.2004555} {\path{doi:10.1080/1461670X.2021.2004555}}.

\bibitem[KM13]{kosara2013storytelling}
\textsc{Kosara R., Mackinlay J.}:
\newblock Storytelling: The next step for visualization.
\newblock \emph{Computer 46}, 5 (2013), 44--50.
\newblock \href {https://doi.org/10.1109/MC.2013.36} {\path{doi:10.1109/MC.2013.36}}.

\bibitem[KMH21]{kim2021design}
\textsc{Kim H., Moritz D., Hullman J.}:
\newblock Design patterns and trade-offs in responsive visualization for communication.
\newblock \emph{Computer Graphics Forum 40}, 3 (2021), 459--470.
\newblock \href {https://doi.org/10.1111/cgf.14321} {\path{doi:10.1111/cgf.14321}}.

\bibitem[LGS{\etalchar{*}}22]{lo2022misinformed}
\textsc{Lo L. Y.-H., Gupta A., Shigyo K., Wu A., Bertini E., Qu H.}:
\newblock Misinformed by visualization: What do we learn from misinformative visualizations?
\newblock \emph{Computer Graphics Forum 41}, 3 (2022), 515--525.
\newblock \href {https://doi.org/10.1111/cgf.14559} {\path{doi:10.1111/cgf.14559}}.

\bibitem[LL24]{lan2024came}
\textsc{Lan X., Liu Y.}:
\newblock "{I} came across a junk": Understanding design flaws of data visualization from the public's perspective.
\newblock \emph{IEEE Transactions on Visualization and Computer Graphics} (2024).
\newblock \url{https://arxiv.org/abs/2407.11497}.

\bibitem[LRIC15]{lee2015more}
\textsc{Lee B., Riche N.~H., Isenberg P., Carpendale S.}:
\newblock More than telling a story: Transforming data into visually shared stories.
\newblock \emph{IEEE Computer Graphics and Applications 35}, 5 (2015), 84--90.
\newblock \href {https://doi.org/10.1109/MCG.2015.99} {\path{doi:10.1109/MCG.2015.99}}.

\bibitem[LWC24]{lan2023affective}
\textsc{Lan X., Wu Y., Cao N.}:
\newblock Affective visualization design: Leveraging the emotional impact of data.
\newblock \emph{IEEE Transactions on Visualization and Computer Graphics 30}, 1 (2024), 1--11.
\newblock \href {https://doi.org/10.1109/TVCG.2023.3327385} {\path{doi:10.1109/TVCG.2023.3327385}}.

\bibitem[LXC21]{lan2021understanding}
\textsc{Lan X., Xu X., Cao N.}:
\newblock Understanding narrative linearity for telling expressive time-oriented stories.
\newblock In \emph{Proceedings of the CHI Conference on Human Factors in Computing Systems} (2021), pp.~1--13.
\newblock \href {https://doi.org/10.1145/3411764.3445344} {\path{doi:10.1145/3411764.3445344}}.

\bibitem[Mar03]{martin2003s}
\textsc{Martin M.~A.}:
\newblock “it's like… you know”: The use of analogies and heuristics in teaching introductory statistical methods.
\newblock \emph{Journal of Statistics Education 11}, 2 (2003), 1--28.
\newblock \href {https://doi.org/10.1080/10691898.2003.11910705} {\path{doi:10.1080/10691898.2003.11910705}}.

\bibitem[Mun14]{munzner2014visualization}
\textsc{Munzner T.}:
\newblock \emph{Visualization analysis and design}.
\newblock CRC press, 2014.
\newblock \href {https://doi.org/10.1201/b17511} {\path{doi:10.1201/b17511}}.

\bibitem[Mut19]{mutsvairo2019challenges}
\textsc{Mutsvairo B.}:
\newblock Challenges facing development of data journalism in non-western societies.
\newblock \emph{Digital Journalism 7}, 9 (2019), 1289--1294.
\newblock \href {https://doi.org/10.1080/21670811.2019.1691927} {\path{doi:10.1080/21670811.2019.1691927}}.

\bibitem[NAK16]{nusrat2016evaluating}
\textsc{Nusrat S., Alam M.~J., Kobourov S.}:
\newblock Evaluating cartogram effectiveness.
\newblock \emph{IEEE Transactions on Visualization and Computer Graphics 24}, 2 (2016), 1077--1090.
\newblock \href {https://doi.org/10.1109/TVCG.2016.2642109} {\path{doi:10.1109/TVCG.2016.2642109}}.

\bibitem[Ngu]{dj_syllabuses}
\textsc{Nguyen D.}:
\newblock Computer-assisted reporting and data journalism syllabuses.
\newblock \url{https://github.com/dannguyen/journalism-syllabi}.
\newblock Last Access: 2025-01-05.

\bibitem[{Nor}]{ne}
\textsc{{Northeastern University}}:
\newblock {Journalism and Data Science, BS}.
\newblock \url{https://camd.northeastern.edu/programs/data-science-journalism-bs/}.
\newblock Last Access: 2025-04-01.

\bibitem[Par15]{parasie2015data}
\textsc{Parasie S.}:
\newblock Data-driven revelation? epistemological tensions in investigative journalism in the age of “big data”.
\newblock \emph{Digital Journalism 3}, 3 (2015), 364--380.
\newblock \href {https://doi.org/10.1080/21670811.2014.976408} {\path{doi:10.1080/21670811.2014.976408}}.

\bibitem[PC15]{plaue2015data}
\textsc{Plaue C., Cook L.~R.}:
\newblock Data journalism: Lessons learned while designing an interdisciplinary service course.
\newblock In \emph{Proceedings of the ACM Technical Symposium on Computer Science Education} (2015), pp.~126--131.
\newblock \href {https://doi.org/10.1145/2676723.2677263} {\path{doi:10.1145/2676723.2677263}}.

\bibitem[RL19]{rees2019survey}
\textsc{Rees D., Laramee R.~S.}:
\newblock A survey of information visualization books.
\newblock \emph{Computer Graphics Forum 38}, 1 (2019), 610--646.
\newblock \href {https://doi.org/10.1111/cgf.13595} {\path{doi:10.1111/cgf.13595}}.

\bibitem[Rog13]{rogers2013facts}
\textsc{Rogers S.}:
\newblock \emph{Facts are sacred: The power of data}.
\newblock Faber \& Faber, 2013.

\bibitem[SA09]{schudson2009objectivity}
\textsc{Schudson M., Anderson C.}:
\newblock Objectivity, professionalism, and truth seeking in journalism.
\newblock In \emph{The handbook of journalism studies}. Routledge, 2009, pp.~108--121.
\newblock \href {https://doi.org/10.4324/9780203877685} {\path{doi:10.4324/9780203877685}}.

\bibitem[SDSE{\etalchar{*}}16]{splendore2016educational}
\textsc{Splendore S., Di~Salvo P., Eberwein T., Groenhart H., Kus M., Porlezza C.}:
\newblock Educational strategies in data journalism: A comparative study of six european countries.
\newblock \emph{Journalism 17}, 1 (2016), 138--152.
\newblock \href {https://doi.org/10.1177/1464884915612683} {\path{doi:10.1177/1464884915612683}}.

\bibitem[SH10]{segel2010narrative}
\textsc{Segel E., Heer J.}:
\newblock Narrative visualization: Telling stories with data.
\newblock \emph{IEEE Transactions on Visualization and Computer Graphics 16}, 6 (2010), 1139--1148.
\newblock \href {https://doi.org/10.1109/TVCG.2010.179} {\path{doi:10.1109/TVCG.2010.179}}.

\bibitem[SH23]{stalph2023exploring}
\textsc{Stalph F., Heravi B.}:
\newblock Exploring data visualisations: an analytical framework based on dimensional components of data artefacts in journalism.
\newblock \emph{Digital Journalism 11}, 9 (2023), 1641--1663.
\newblock \href {https://doi.org/10.1080/21670811.2021.1957965} {\path{doi:10.1080/21670811.2021.1957965}}.

\bibitem[SLL{\etalchar{*}}21]{shi2021communicating}
\textsc{Shi Y., Lan X., Li J., Li Z., Cao N.}:
\newblock Communicating with motion: A design space for animated visual narratives in data videos.
\newblock In \emph{Proceedings of the CHI Conference on Human Factors in Computing Systems} (2021), pp.~1--13.
\newblock \href {https://doi.org/10.1145/3411764.3445337} {\path{doi:10.1145/3411764.3445337}}.

\bibitem[SLRS18]{stolper2018data}
\textsc{Stolper C.~D., Lee B., Riche N.~H., Stasko J.}:
\newblock Data-driven storytelling techniques: Analysis of a curated collection of visual stories.
\newblock In \emph{Data-driven storytelling}. AK Peters/CRC Press, 2018, pp.~85--105.
\newblock \href {https://doi.org/10.1201/9781315281575} {\path{doi:10.1201/9781315281575}}.

\bibitem[{Soc}14]{spj}
\textsc{{Society of Professional Journalists}}:
\newblock {SPJ} code of ethics.
\newblock \url{https://www.spj.org/ethicscode.asp}, 2014.

\bibitem[{The}a]{investigative_j}
\textsc{{The Centre for Investigative Journalism}}:
\newblock Data journalism.
\newblock \url{https://tcij.org/bespoke-training/data-journalism/}.
\newblock Last Access: 2025-04-01.

\bibitem[{The}b]{wiki_guardian}
\textsc{{The Guardian}}:
\newblock Datablog + wikileaks.
\newblock \url{ https://www.theguardian.com/news/datablog+media/wikileaks}.
\newblock Last Access: 2025-01-05.

\bibitem[{The}c]{wiki_nyt}
\textsc{{The New York Times}}:
\newblock The war logs.
\newblock \url{ https://archive.nytimes.com/www.nytimes.com/interactive/world/war-logs.html\#nytint-afghan}.
\newblock Last Access: 2025-01-05.

\bibitem[TRLL16]{treadwell2016numbers}
\textsc{Treadwell G., Ross T., Lee A., Lowenstein J.~K.}:
\newblock A numbers game: Two case studies in teaching data journalism.
\newblock \emph{Journalism \& Mass Communication Educator 71}, 3 (2016), 297--308.
\newblock \href {https://doi.org/10.1177/1077695816665215} {\path{doi:10.1177/1077695816665215}}.

\bibitem[Ush16]{usher2016interactive}
\textsc{Usher N.}:
\newblock \emph{Interactive journalism: Hackers, data, and code}.
\newblock University of Illinois Press, 2016.
\newblock \href {https://doi.org/10.5406/illinois/9780252040511.001.0001} {\path{doi:10.5406/illinois/9780252040511.001.0001}}.

\bibitem[vis]{vislie}
{VisLies}.
\newblock \url{https://www.vislies.org}.
\newblock Last Access: 2025-04-01.

\bibitem[VJM09]{vallance2009computer}
\textsc{Vallance-Jones F., McKie D.}:
\newblock \emph{Computer-assisted reporting: A comprehensive primer}.
\newblock Oxford University Press, USA, 2009.

\bibitem[WEK18]{weber2018data}
\textsc{Weber W., Engebretsen M., Kennedy H.}:
\newblock Data stories: Rethinking journalistic storytelling in the context of data journalism.
\newblock \emph{Studies in Communication Sciences 2018}, 1 (2018), 191--206.
\newblock \href {https://doi.org/10.24434/j.scoms.2018.01.013} {\path{doi:10.24434/j.scoms.2018.01.013}}.

\bibitem[WN23]{wright2023development}
\textsc{Wright S., Nolan D.}:
\newblock The development of data journalism in china: influences, motivations and practice.
\newblock \emph{Digital Journalism 11}, 9 (2023), 1664--1681.
\newblock \href {https://doi.org/10.1080/21670811.2021.1927779} {\path{doi:10.1080/21670811.2021.1927779}}.

\bibitem[Won13]{wong2013wall}
\textsc{Wong D.~M.}:
\newblock \emph{The Wall Street Journal guide to information graphics: The dos and don'ts of presenting data, facts, and figures}.
\newblock WW Norton \& Company, 2013.

\bibitem[YXL{\etalchar{*}}22]{yang2021design}
\textsc{Yang L., Xu X., Lan X., Liu Z., Guo S., Shi Y., Qu H., Cao N.}:
\newblock A design space for applying the freytag's pyramid structure to data stories.
\newblock \emph{IEEE Transactions on Visualization and Computer Graphics 28}, 1 (2022), 922--932.
\newblock \href {https://doi.org/10.1109/TVCG.2021.3114774} {\path{doi:10.1109/TVCG.2021.3114774}}.

\bibitem[ZC14]{zion2014ethics}
\textsc{Zion L., Craig D.}:
\newblock \emph{Ethics for Digital Journalists}.
\newblock Routledge, 2014.
\newblock \href {https://doi.org/10.4324/9780203702567} {\path{doi:10.4324/9780203702567}}.

\end{thebibliography}
